# Enhancing Students' Understanding of Scientific Equipment: Smartphones in the Laboratory

Colleen L. Countryman and M. A. Paesler

North Carolina State University, 2401 Stinson Dr., Campus Box 8202, Raleigh, NC 27695; colleen_countryman@ncsu.edu, paesler@ncsu.edu

One of the oft-cited qualities[1,2] sought after in a potential future engineering employee is an analytical mind that is "continually examining things." In one sense this examination is discouraged in an instructional laboratory employing opaque black box data collection devices. Smartphones can be used to gather such data but also – by their very nature—have the capability of removing the veil of opacity so as to allow students to examine their operation at a very fundamental level. By taking advantage of a smartphone's visualization capabilities one can elucidate core aspects of its operation equally[3] or perhaps better than[4] a mechanical model might. Through purposeful app and curricular design based on student feedback, we avoid this commonly perceived[5,6] pitfall of electronic devices and encourage student examination of one such device, the student's own smartphone.

## The Potential Danger of "Black Boxes"

The collection and interpretation of data in instructional laboratories provide pedagogical value to a physics curriculum. Learning in such laboratory activities can be enhanced when students understand the operation of data collection equipment. In a 1993 article in *The Physics Teacher*, former AAPT president Arnold Arons identified this value when he underscored the importance of "subjecting a piece of equipment to close examination in context, figuring out how it works and how it might be used (rather than simply being *told* how it works and what it is supposed to do)."[7] Other educational researchers have also pointed out the necessity of modeling measurement tools as a "prerequisite for engaging in experimental design."[8]

The goals of a typical introductory lab are often ambitious including mastery in content, experimental design, data analysis, modeling and communication. An additional important component, however, is the development of practical laboratory skills. To this end, instructors are often warned of the hazards of using "black boxes" – that is, data collection instruments with opaque housings and enigmatic inner-workings[5]. Resnick et al.[6] decry modern scientific instruments used in teaching laboratories saying "their inner workings are often hidden and thus poorly understood by their users." They further suggest that "digital electronics and computational technologies have accelerated this trend, filling science laboratories and classrooms with ever more opaque black boxes."

Students and instructors alike are realizing the potential for capturing motion data of a smartphone housed in its internal accelerometer and gyroscope. In BYOD (Bring Your Own Device) labs that rely upon students' technologies, educational researchers have alluded to the possibility of these devices contributing to the instrument opacity problem prevalent in instructional labs[9,10,11]. For example, students often struggle to understand why a smartphone accelerometer would read a nonzero acceleration when the phone is at rest. Previous work has identified possible curricular changes to address these concerns[12]. We now address these same pedagogical concerns through careful design of a mobile app and corresponding laboratory curriculum.

## Preliminary Study

Prior to the development of a new curriculum and construction of our own app, a few laboratory sections composed primarily of engineering majors were modified to allow for use of smartphones in lieu of conventional laboratory equipment. The labs were absent of any direct instruction regarding the accelerometer's inner-workings. After asking students to use their smartphone accelerometers for data collection in some simple kinematic experiments, we asked them to describe how the accelerometer worked. To encourage the use of everyday parlance with sophisticated responses, we specifically asked them the following:

"Suppose your 12-year-old brother came to visit your lab. At the end of the lab he asks, 'How did the phone measure acceleration?' How would you answer his question? Diagrams are welcome!"

When students used some existing free accelerometer apps[13,14] with a traditional curriculum adapted minimally for the inclusion of this technology, students were not able to describe the inner-workings of the accelerometer. A small sample of $n = 10$ students presented the following explanations, including the pictorial explanation in Figure 1:

- The gyroscope in the phone can measure things, such as acceleration, in three directions dimensions $(x, y, z)$. When the phone experiences accelerations in either three, the gyroscope can calculate the acceleration by the amount of change in work it experiences to 'stabilize' itself.
- The phone has a gyroscope which allows data to be taken in 3D. Using Bluetooth and wireless signals the phone is also able to measure speed (velocity) and acceleration based on location.
- Magic.
- Because an app on the phone allows it to use an accelerometer.
- "There is an accelerometer in this phone that can measure how quickly it moves around." I'd start up the phone's accelerometer and explain the graph.
- I would be able to explain the general physics principles behind certain concepts however not specific how the phone was able to measure the data.
- It has a device in it that measures the force of gravity in certain directions.
- I have no idea. I just know how to read it.
- The sensor in the phone, that has the GPS, the sensor for making the screen sleep when talking.

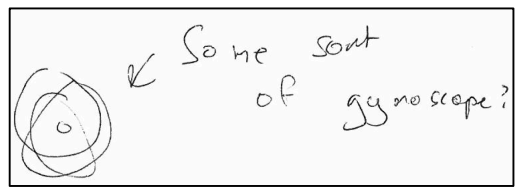

*Figure 1. One student's possible explanation for how the phone's internal accelerometer collects data. The text reads "Some sort of gyroscope?"*

These responses lack detail or physical explanations for what is happening inside the accelerometer. Indeed, many of the explanations relied upon technological terms that related little to the topic at hand. For example, GPS, wifi, and Bluetooth have nothing to do with how the accelerometer collects data, and the gyroscope is a distinct sensor within the phone and does not contribute to the production of the raw accelerometer data.

To study student reactions more deeply and to guide us in our efforts to avoid the black box pitfalls, we requested student feedback with surveys and interviews. These led us to a series of challenges in our attempts to move forward. Most pertinent to the issues at hand was the observation that—as expected—students had very little understanding of the data collection mechanism.

## Development of the MyTech App and Corresponding Curriculum

The challenge revealed in our preliminary work guided the development of our own app better suited to deal with inherent black box problems. Dubbed "MyTech," the app and its usage incorporate several important features[15].

Perhaps most importantly, before students use the app to perform kinematic experiments, they undertake an initial exploratory laboratory exercise designed to introduce them to the operation of the smartphone as a kinematic data-taking device[12]. As Ronald Thornton indicates, any "well designed measurement tool can help," but inquiry-based "accompanying curricula" are necessary to optimize students' understanding[16]. Our exercise familiarizes students with the orientation of the three Cartesian axes on their smartphones and the actual operation of the three internal accelerometers. To aid in accomplishing these goals, several features are incorporated into the MyTech app itself.

First, to eliminate the black box pedagogical opacity, the app includes a visualization of a mass and spring model of a one-dimensional accelerometer. As an internal accelerometer is indeed composed of a mass attached loosely to the phone with a few flexible silica arms, the ball and spring representation is an appropriate model of the accelerometer's internal action. Zacharia and Constantinou demonstrated that virtual manipulatives can be as effective as physical ones in their impact on students' conceptual understanding. Furthermore, in a similar study, Finkelstein, et al. illustrated the improved conceptual understanding of students that use their simulated equipment (specifically, PhETs, or Physics Education Technologies) as compared to those that use real equipment in lab settings. In the first exploratory lab, students are encouraged to examine the ball and spring visualization (seen in the left image of Figure 2) and to describe their observations as they moved the phone up and down. Utilizing common active learning methods,[17,18,19] students are also encouraged to predict the accelerometer graphs they might obtain from a series of hypothetical experiments.

Secondly, a series of help articles on the app (seen in the right image of Figure 2) addresses common troubleshooting issues and guides students' understanding of the device. Students are not explicitly required to view the help articles, but tracking of usage suggests that approximately 13% of the students explored the help menu. The most popular of these help articles is titled "What are the accelerometer axes?"

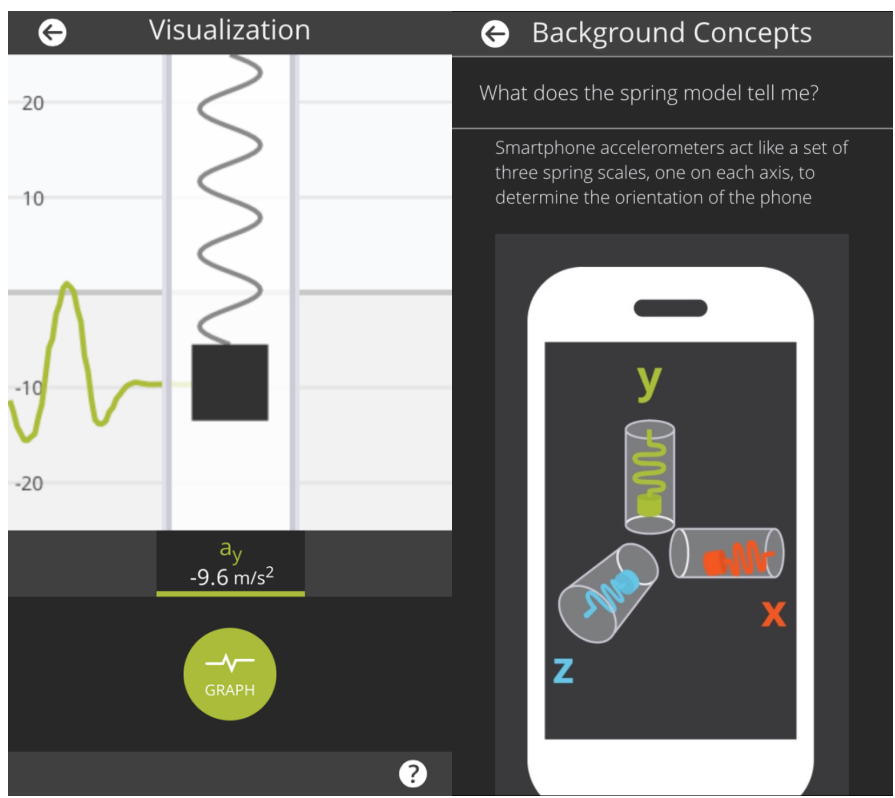

*Figure 2. In the "visualization" mode of the NCSU MyTech app (as shown on the left), students can explore how a simple spring model of a one-dimensional accelerometer reacts to motion of the smartphone. The help articles (an example of which is shown on the right) address introductory physics students' common questions. One help article provides further detail about how accelerometers work.*

## Post-MyTech App Study

In the second round of the study, a new group of students were asked the same question (about explaining the accelerometer's inner-workings to their 12-year-old brother), after exposure to the MyTech app and associated curriculum. The responses ($n = 41$) were different than the previous responses. A sampling are presented below and in Figure 3, corrected only for grammar:

- I would tell him that there are essentially three sensors corresponding to a vertical, and two horizontal axes. These sensors measure force. Pretend there's a ball inside of a box. When you move the box, the ball applies a force on the sides of a box, and this force is measured to determine in which direction and how fast this box is moving.
- I would tell my brother about the spring mechanism inside of my phone and explain to him how it works to find the acceleration due to gravity as the phone is at rest. I would then show him the visualization spring on the different axes as I rotate the phone and it should help to show him how the accelerometer works.

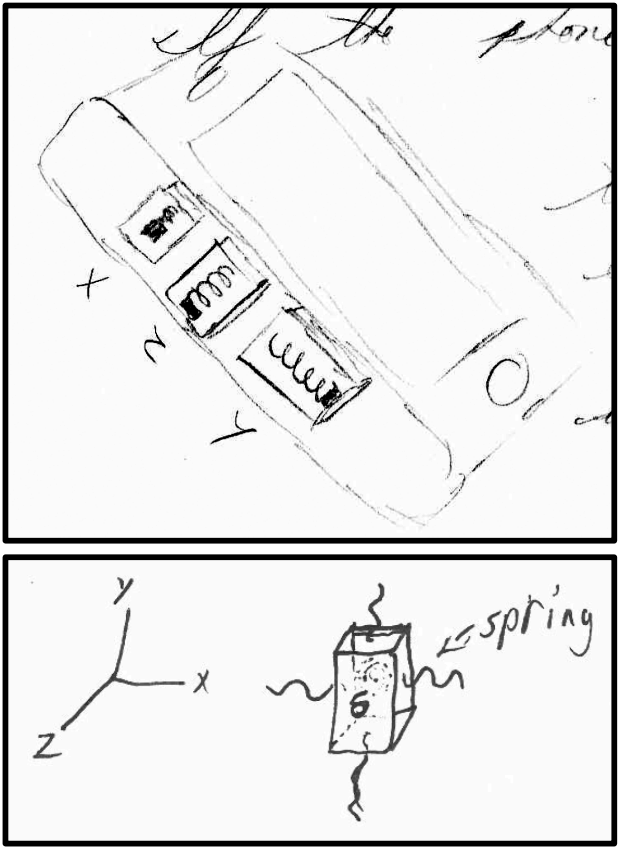

*Figure 3. After using the NCSU MyTech app, students provided more accurate explanations with greater detail of how the accelerometer works. Above, a student shows a separate spring model of an accelerometer along the three dimensions of the smartphone. Below, a student shows a proof mass connected by springs to each dimension of the smartphone.*

Additionally, students' self-reported survey data indicates the value of the accelerometer visualization and help articles, as illustrated in Figure 4. In fact, most of the class either agreed or strongly agreed with the fact that the spring visualization aided their understanding of how a smartphone accelerometer works. (Only a few students disagreed, and none strongly disagreed.) Similarly, many of the students claimed that the help articles resolved their questions or challenges, while some disagreed, and none strongly disagreed.)

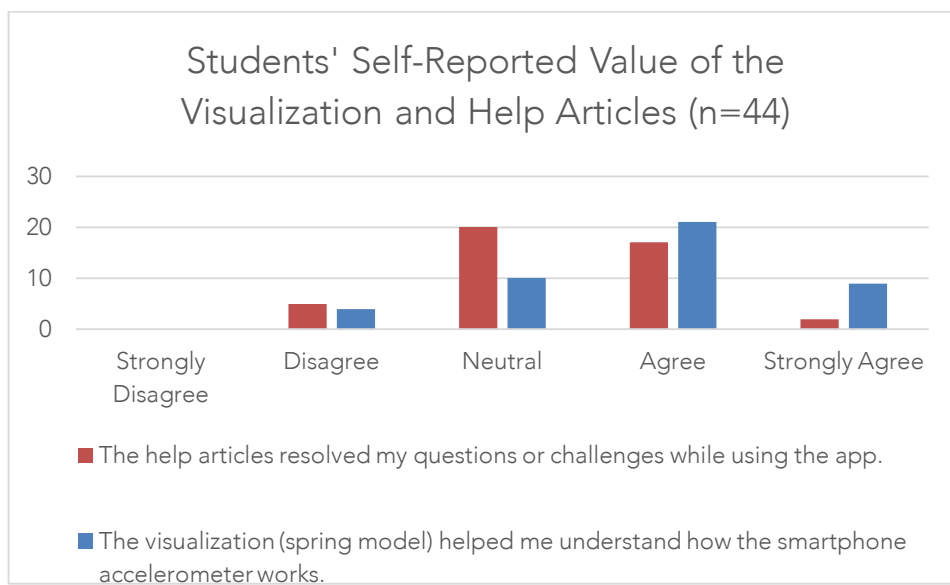

*Figure 4. Students believe that the help articles and spring model visualization of the accelerometer allowed them to resolve questions and improve their understanding of the smartphone accelerometer, respectively.*

## Conclusions

Students are capable of understanding how smartphone data collection mechanisms work because of the increased transparency of the equipment in our introductory physics laboratories. In fact, with a directed app and lab curriculum, students in our courses exceed the minimum recommendations for introductory students[20]. They are not only capable of using the accelerometer to take data, but they can also explain in some detail how it works.

## Acknowledgments

The work was conducted with support from NSF (grant #1245832, under the Directorate for Education and Human Resources) and the North Carolina State University DELTA Exploratory Grant. The authors are especially grateful to Robert Beichner, Brad Mehlenbacher and the entire NCSU PER Group for helpful discussions. Additionally, we recognize David Tredwell, Yan Shen, Chrissie Van Hoever, and Samantha McCuen at NCSU DELTA for their expertise in developing and assessing the impact of the MyTech app.